\documentclass[a4paper,aps,prl,superscriptaddress,floatfix,nofootinbib,twocolumn,bibtex]{revtex4}

\usepackage{bbold}
\usepackage{bbm}
\usepackage[pdftex]{graphicx}
\usepackage{latexsym,amsmath,verbatim}
\usepackage{color}
\usepackage{rotating}
\usepackage{verbatim}
\usepackage{multirow}
\usepackage[english]{babel}
\usepackage{comment}
\usepackage{ulem}

\begin{document}

\title{Beyond the locally tree-like approximation for percolation on
  real networks}

\author{Filippo Radicchi}
\affiliation{Center for Complex Networks and Systems Research, School of Informatics and Computing, Indiana University, Bloomington, USA}
\email{filiradi@indiana.edu}

\author{Claudio Castellano}
\affiliation{Istituto dei Sistemi Complessi (ISC-CNR), Roma, Italy, and
Dipartimento di Fisica, Sapienza Universit\`a di Roma, Roma, Italy}

\begin{abstract}
  Theoretical attempts proposed so far to describe ordinary
  percolation processes on real-world networks rely on the
  locally tree-like ansatz.  Such an approximation, however, holds
  only to a limited extent, as real graphs are often characterized by
  high frequencies of short loops. We present here a
  theoretical framework able to overcome such a limitation
  for the case of site percolation. 
  Our method is based on a
  message passing algorithm that discounts redundant paths along
  triangles in the graph. We systematically test the approach on $98$
  real-world graphs and on synthetic networks.  We find excellent
  accuracy in the prediction of the whole percolation diagram, with
  significant improvement with respect to the prediction
  obtained under the locally tree-like approximation. 
  Residual discrepancies between theory and simulations do not depend on
  clustering and can be attributed to the presence of loops longer than 
  three edges. We present also a method to account for clustering in
  bond percolation, 
  but the improvement with respect to the method based on 
  the tree-like approximation is much less apparent.
\end{abstract}

\maketitle

Percolation processes are often used to study resilience properties of
real networks~\cite{albert2000error, cohen2000resilience,
  callaway2000network}, and play a fundamental role in the
understanding of spreading phenomena in real
systems~\cite{pastor2001epidemic, newman2002spread}.  Percolation has
been intensely studied in a multitude of network
models~\cite{dorogovtsev2008critical, dorogovtsev2010lectures,
  newman2010networks}, including sparse tree-like graphs
~\cite{cohen2000resilience, callaway2000network,
  cohen2002percolation}, as well as generative models for random
networks with triangles, cliques or arbitrary
subgraphs~\cite{Serrano06,Serrano06b,Gleeson09,
  newman2009random,Miller09,Gleeson10}.
These studies shed light on fundamental physical 
mechanisms of percolation processes on complex network topologies, but their
importance in the analysis of percolation on real-world graphs
is limited, as the topology of individual real networks often markedly differs
from the one of random network ensembles.  Recent works have attempted to
overcome such a serious limitation.  Karrer {\it et al.}
formulated a novel method which takes as input the detailed
topological structure of a given network to predict the value of the
percolation strength (and other macroscopic observables) as a function
of the bond occupation probability~\cite{PhysRevLett.113.208702}.  In
particular, they demonstrated that the bond percolation threshold of a
given network is bounded from below by the leading eigenvalue of its
non-backtracking matrix~\cite{hashimoto1989zeta}.  An approach based
on the same rationale was also used by Hamilton and Pryadko to study
site percolation in isolated networks~\cite{PhysRevLett.113.208701},
and by Radicchi in the analysis of bond and site percolation models in
interdependent networks~\cite{radicchi2015percolation}.  These methods
still suffer from a fundamental limitation: they are based on the
locally tree-like approximation~\cite{dorogovtsev2008critical,
  dorogovtsev2010lectures, newman2010networks}, and as such they are
potentially
not reliable for networks with nonnegligible density of
triangles, or short loops in general~\cite{radicchi2014predicting, PhysRevE.91.052807}.

In this paper, we make a step forward, by generalizing the approach
developed in~\cite{PhysRevLett.113.208701, PhysRevLett.113.208702} to
clustered networks. Through a systematic analysis of about one hundred
real-world networks as well as clustered synthetic ones, we
demonstrate that our 
framework provides excellent prediction of the whole phase diagram 
for the site percolation model. Furthermore, we present an approach
improving also the prediction of the bond percolation phase diagram
(though in a less satisfactory way) and understand the origin
of the differences between the two cases.

We start our analysis from the site percolation model. We assume that
the structure of a network with $N$ nodes and $E$ edges is given by a
one-zero adjacency matrix $A$ (i.e., the generic element $A_{i,j} = 1$
if vertices $i$ and $j$ are connected, whereas $A_{i, j} = 0$
otherwise). We further assume that the network is composed of a single
connected component.  In the ordinary site percolation model, each
node is active or occupied with probability $p$.  Two active nodes
belong to the same cluster if there exists at least a path, passing
only through active nodes, that connects them.  For $p=0$, no nodes
are active so that there are no clusters. For $p=1$, all nodes are
active and belong to a single cluster of size $N$. As $p$ varies, the
network undergoes a structural phase transition, at the percolation
threshold $p_c$, corresponding to the appearance of an extensive cluster.  
The transition can be monitored through the
so-called percolation strength $S_\infty$, defined as the relative
size of the largest cluster with
respect to the size of the network.  For $p=0$, $S_\infty = 0$; for
$p=1$, $S_\infty = 1$. The goal of the following approach is to
estimate the expected value of $S_\infty$ over an infinite number of
realizations of the percolation model for any given value of $p$.  The
probability $s_i$ that node $i$ belongs to the largest cluster can be
described by the equation
\begin{equation}
s_i = p [1 - \prod_{j \in \mathcal{N}_i} (1 - t_{i \to j}) ] \; ,
\label{eq:site1}
\end{equation}
where $\mathcal{N}_i$ is the set of neighbors of node $i$ and $t_{i \to
  j}$ quantifies the probability that following the edge $(i, j)$, in
the direction $i \to j$, we find a node belonging to the largest
cluster. The quantity $t_{i \to j}$ can be interpreted as a
``message'' passed from node $j$ to vertex $i$ about belonging to
the largest cluster. Eq.~(\ref{eq:site1}) essentially states that the
probability that node $i$ is part of the largest cluster equals the
product of the probabilities that (i) node $i$ is active and (ii) at
least one of its neighbors is in the largest cluster. For consistency,
the probability $t_{i \to j}$ is described by the equation
\begin{equation}
t_{i \to j} = p [1 - \prod_{k \in \mathcal{Q}_{i \to j}} (1 - t_{j \to k}) ] \; .
\label{eq:site2}
\end{equation}
The explanation of this equation is similar to the previous one. The
only difference here is that the product does not run necessarily over
all the neighbors of node $j$, but only on the elements of the set
$\mathcal{Q}_{i \to j}$. We note that, while Eq.~(\ref{eq:site2}) is
in principle defined for every pair of node indices $i \to j$, only
pairs of nodes connected by an edge play a role in
Eq.~(\ref{eq:site1}). We have therefore $2E$ equations of the
type~(\ref{eq:site2}) that can be solved by iteration. The solutions
of these equations
are then plugged into the set of
$N$ Eqs.~(\ref{eq:site1}) to
determine the value of every $s_i$. Finally, the
percolation strength is computed as
\begin{equation}
S_\infty = \frac{1}{N} \sum_i s_i \; .
\label{eq:site}
\end{equation}
Since the entire operation can be repeated for any value of the
occupation probability $p$, Eqs.~(\ref{eq:site1}), ~(\ref{eq:site2})
and~(\ref{eq:site}) allow to draw the entire phase diagram for a given
network.  A linear expansion of the system of Eqs.~(\ref{eq:site2})
can be used to obtain an eigenvalue/eigenvector equation of the type
$\vec{t} = p \, G \, \vec{t}$, where $\vec{t}$ is a vector with $2E$
components, and $G$ is a $2E \times 2E$ one-zero matrix. 
A non trivial solution exists only if $1/p$ is an eigenvalue of the
operator $G$.
Thus the inverse of the largest eigenvalue of $G$
(which is real according to the Perron-Frobenius theorem)
is the percolation threshold $p_c$ of the network.

\begin{figure}[!htb]
  \begin{center}
    \includegraphics[width=0.47\textwidth]{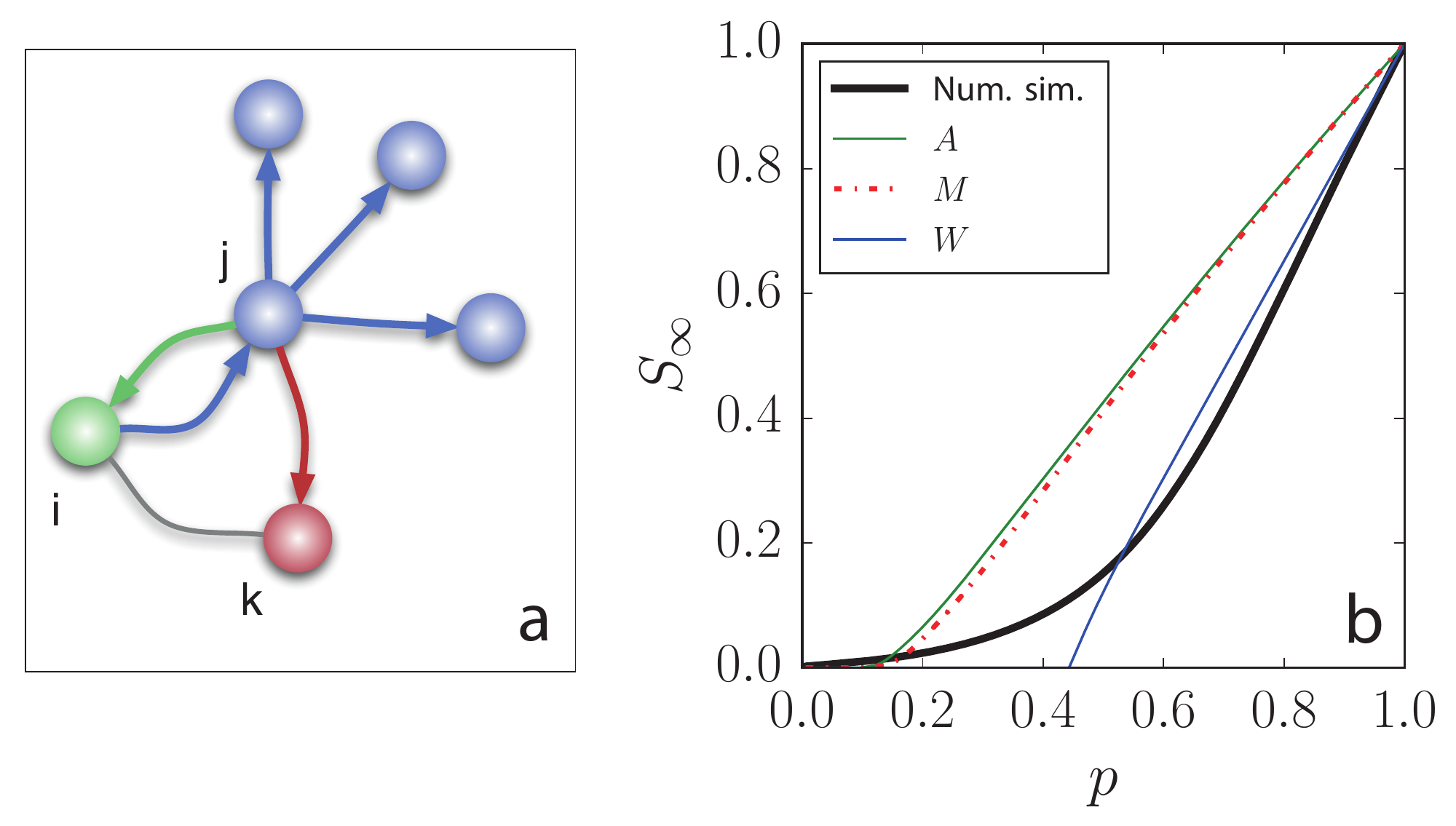}
  \end{center}
  \caption{(a) Illustration of the different ways of defining
    $\mathcal{Q}_{i \to j}$ in Eq.~(\ref{eq:site2}). The $A$-based
    approximation is obtained by setting
    $\mathcal{Q}_{i \to j} = \mathcal{N}_j$, so all edges departing from
    $j$ are included in the equation.  If backtracking walks are
    excluded, that is $\mathcal{Q}_{i \to j} = \mathcal{N}_j \setminus
    \{i\}$, the message passing equation will not include the green
    edge. we refer to this approximation as
    the $M$-based approximation.  If $\mathcal{Q}_{i \to j} = \mathcal{N}_j \setminus [\{i\}
      \cup (\mathcal{N}_j \cap \mathcal{N}_i)]$ so one can only move
    away from node $i$, also walking to node $k$ is avoided and the
    only terms appearing in the equation are given by those
    corresponding to the blue arrows. This corresponds
    to the $W$-based approximation.  (b) Phase diagram for the site
    percolation model applied to co-authorship graph among network
    scientists~\cite{newman2006finding}. The black line denotes the
    results of numerical simulations of the model. The other curves
    represent results obtained through the numerical solution of
    Eqs.~(\ref{eq:site1}),~(\ref{eq:site2}) and~(\ref{eq:site}) with
    different definitions of $\mathcal{Q}_{i \to j}$.
  }
\label{fig1}
\end{figure}

The form of $G$ depends on the definition of the set $\mathcal{Q}_{i
  \to j}$ in Eq.~(\ref{eq:site2}), which is crucial for the
effectiveness of the entire approach.
We illustrate here three different, and increasingly accurate,
approximations (see Fig.~\ref{fig1}a).  In the first
approximation, we set $\mathcal{Q}_{i \to j} = \mathcal{N}_j$.  Such a
choice makes Eq.~(\ref{eq:site2}) identical to Eq.~(\ref{eq:site1}),
so that $t_{i \to j} = s_j$. The generic
element of the matrix $G$ is $G_{i \to j, \ell \to k} = \delta_{j, \ell}$, 
with $\delta_{x, y} $ the Kronecker symbol. 
This matrix has the same eigenvalues of the adjacency
matrix~\cite{PhysRevLett.113.208702}. Hence 
the percolation threshold under this approximation is given
by the inverse of the  leading eigenvalue of the 
adjacency matrix~\cite{bollobas2010percolation}.
We refer to it
as the adjacency-matrix-based or, in short, $A$-based approximation. 
In this approximation,
the variable $t_{j \to i}$ is on the r.h.s. of Eq.~(\ref{eq:site2}),
so that $t_{i \to j}$ grows as 
$t_{j \to i}$ increases.  In turn, the value of $t_{j \to i}$ is also
increased by the growth of $t_{i \to j}$.  The possibility for a
message to pass back and forth on the same edge causes an ``echo
chamber'' effect in the
equations that leads to an overestimation of the correct values of the
variables $t$ and hence of the percolation strength. 
To suppress this effect, a more precise approximation prescribes 
$\mathcal{Q}_{i \to j} = \mathcal{N}_j \setminus \{i\}$.  
The motivation of this choice is simple: the exclusion of
vertex $i$ from the product on the r.h.s. of Eq.~(\ref{eq:site2}) does
not allow for backtracking messages, and the variable
$t_{j \to i}$ does not appear anymore
on the r.h.s. of Eq.~(\ref{eq:site2}). Under this
approximation, $G$ coincides with $M$, the non-backtracking matrix of
the graph~\cite{PhysRevLett.113.208702}, whose generic
element is
\begin{equation}
M_{i \to j, \ell \to k} = \delta_{j, \ell} (1 - \delta_{i, k}) \; .
\label{eq:nbt}
\end{equation}
The percolation threshold is estimated as the inverse of the principal
eigenvalue of the non-backtracking matrix of the
graph~\cite{PhysRevLett.113.208702, radicchi2015percolation}.  The
$M$-based approximation is exact in networks with locally tree-like
structure. However, if loops are present in the network, echo
chamber effects still persist.  This undesirable effect can be once
more discounted by excluding redundant paths caused by
triangles, that is using the following approximation
\begin{equation}
\mathcal{Q}_{i \to j} = \mathcal{N}_j 
\setminus [ \{i\} \cup (\mathcal{N}_j \cap \mathcal{N}_i)] \; .
\label{second}
\end{equation}
The rationale behind Eq.~(\ref{second}) is again intuitive.  If we are
looking at the
network from vertex $i$, we should disregard the path $i \to
j \to k$ if we already considered the edge $i \to k$, otherwise vertex
$i$ would receive twice the same message from node $k$. The importance
of this correction is apparent in Fig.~\ref{fig1}b, where
the results of simulations for the site percolation model
[see Supplemental Material (SM) for details] are compared
with the numerical solutions of
Eqs.~(\ref{eq:site1}),~(\ref{eq:site2}) and~(\ref{eq:site}) adopting
the three different definitions of $\mathcal{Q}_{i \to j}$ illustrated
above. The network analyzed in Fig.~\ref{fig1}b is a
graph of scientific co-authorships characterized by a very high value
of the clustering coefficient ($C = 0.7412$)~\cite{newman2006finding}.
As in the cases of the first two approximations, also
the last, new approximation allows for the computation of the
percolation threshold through the linearization of
Eqs.~(\ref{eq:site2}). The critical value of the occupation
probability is given by the inverse of the leading eigenvalue of the
matrix $G = W$, defined as
\begin{equation}
  W_{i \to j, \ell \to k} = \delta_{j, \ell} \, (1 - \delta_{i,k})  (1 - A_{i,k}) \; .
  \label{eq:walk}
\end{equation}
The definition of the matrix $W$ is very similar to the one of the
non-backtracking matrix appearing in Eq.~(\ref{eq:nbt}).  The only
difference is the additional term $(1 - A_{i,k})$, that excludes
connections among edges that are part of a triangle. In the matrix
$W$, the directed edges $i \to j$ and $\ell \to k$ are connected only
if $j=\ell$, and node $k$ is at distance two from vertex $i$. 
Mathematical arguments analogous to those presented by Karrer
{\it et al.}~\cite{PhysRevLett.113.208702} (see SM) 
show that the percolation threshold predicted 
using the $W$-based approximation is
always larger than or equal to the one predicted using the $M$-based
method (with the equality sign valid when no triangles are
present), and always smaller than or equal to the
true percolation threshold. Both these inequalities are
validated in all numerical experiments on both real
and synthetic networks. 
For the network of Fig.~\ref{fig1}b, the $A$-based approximation
predicts $p_c^{(A)} = 0.0964$; the approximation based on the $M$ matrix gives
$p_c^{(M)} = 0.1148$; the approximation based on $W$ provides instead
$p_c^{(W)} = 0.4436$. Those predictions compared to the best estimate
$p_c = 0.6300$ from numerical simulations have associated relative
errors respectively equal to $r^{(A)} = 0.8470$, $r^{(M)} = 0.8178$
and $r^{(W)} = 0.2959$.  These correspond to an improvement of roughly
$3\%$ from the $A$-based to the $M$-based approximation, 
and more than $50\%$ from the $M$-based to the $W$-based approximation.
The situation is qualitatively and quantitatively similar in all other
real networks we consider in this study (see SM).  
We can conclude that the inverse of the largest eigenvalue of the matrix
$W$ represents a tighter lower-bound of the true site percolation
threshold than the analogous quantity computed using the $M$ matrix.

\begin{figure}[!htb]
  \begin{center}
    \includegraphics[width=0.47\textwidth]{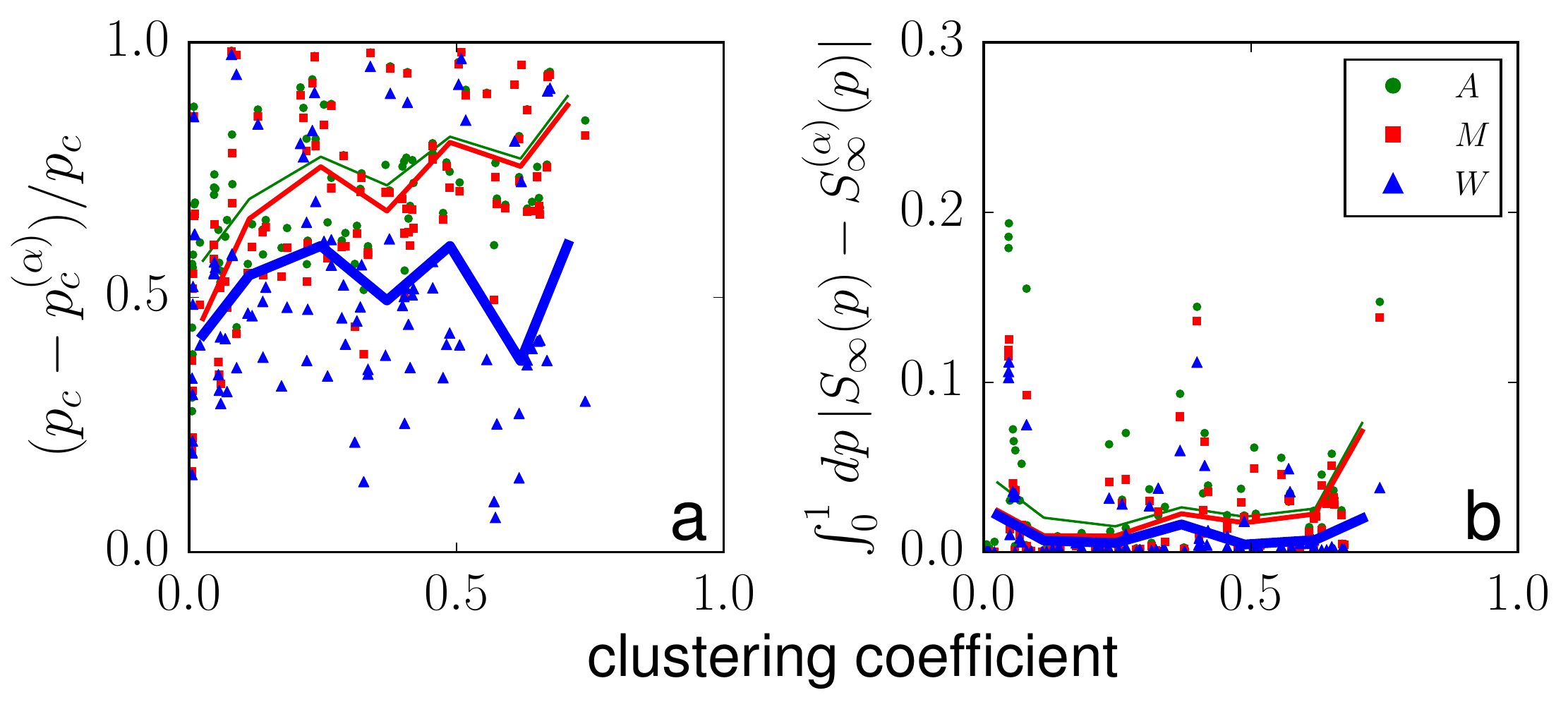}
  \end{center}
  \caption{(a) Relative error associated to the estimation of the site
    percolation threshold for $98$ real networks. For the
    $\alpha$-based approximation, the relative error is measured as
    $(p_c - p_c^{(\alpha)}) / p_c$, with
    $p_c^{(\alpha)}$ best estimate of the percolation threshold for
    the $\alpha$-based approximation [$\alpha = A, M$ or $W$], 
    and $p_c$ value of the
    occupation probability corresponding to the peak of the
    susceptibility. The relative error is plotted against the average
    clustering coefficient of the network. Different colors and
    symbols correspond to different orders of the approximation. Full
    lines indicate average values of the relative error for networks
    with similar values of clustering coefficient.
    They are generated according to the 
    following procedure. We divide the 
    range of possible values of $C$ in seven equally spaced
    bins. We then estimate the average value of the error in each bin,
    and the average value of the cluster coefficient within each bin.
    The lines are obtained connecting these points. 
    (b) Same as in panel a, but for the global error measure 
    $\int_0^1 dp |S_{\infty}(p) - S_{\infty}^{(\alpha)}(p)|$.
  }
\label{fig2}
\end{figure}

The $W$-based approximation is able to reproduce with
impressive accuracy the whole percolation diagram of almost all 
the $98$ real networks we analyzed~\cite{radicchi2015percolation}. 
The only exceptions are spatially embedded networks and a few others,
where the $W$-based approximation greatly outperforms the
other approximations but still differs significantly from the
numerical simulations.
The results of our analysis are summarized in Fig.~\ref{fig2}a, where
relative errors in the estimates of the percolation threshold are
plotted against the average clustering coefficients of the networks. 
\footnote{The rather large values of the errors in Fig.~\ref{fig2}a
are also an effect of the difficulties in the numerical estimate of 
the threshold for small networks. See SM for details.}  We
quantify the performance of the various approximations also in terms
of the global error measure~\cite{Melnik11} 
$\int_0^1 dp |S_{\infty}(p) - S_{\infty}^{(\alpha)}(p)|$, with 
$\alpha = A, M$ or $W$ (Fig.~\ref{fig2}b).
We remark that the discrepancy between the $W$-based
approximation and simulations is essentially independent of the
clustering coefficient $C$. This happens because the $W$-based 
approximation becomes exact in the infinite size limit for 
site percolation on  networks containing only short loops
of length three (see SM), 
such as two important classes of random network models with large 
clustering~\cite{Gleeson09,newman2009random,Miller09,Gleeson10}. 
The residual discrepancies in Fig. 2 depend only on the presence
of longer loops, which do not contribute to the value of $C$.
In the SM we also show that the $W$-based approximation 
can be in principle further improved to account for loops of length 
longer than three, but that a systematic approach becomes
practically unfeasible already for loops of length four.  

Next, we consider ordinary bond percolation on a given
network. In this model, every edge is present or active with
probability $p$. Clusters are formed by nodes connected by at least
one path composed of active edges. The order parameter used to monitor
the percolation transition, from the disconnected configuration at
$p=0$ to the globally connected configuration for $p=1$, is still
given by the relative size of the largest connected cluster, namely
$B_\infty$.  The message passing equations valid for the approximations 
based on the adjacency and on the non-backtracking matrices are identical to those already
written for the site percolation model, with the only difference of a factor
$p$~\cite{RadicchiCastellano15}.  The order parameters are
related by $B_\infty = p^{-1}\, S_\infty$, and the percolation
thresholds predicted by the equations are identical in the two
models~\cite{PhysRevLett.113.208701, PhysRevLett.113.208702,
RadicchiCastellano15}.  Writing an improved approximation able to fully
take into account triangles, such as the $W$-based approximation for site
percolation, is in this case impossible (see SM). 
However, one can still write a similar approach
which improves with respect to the two old methods.
In the bond percolation model, a
triangle is effectively present only if all its edges are
simultaneously active, leading to the following self-consistent equations
\begin{equation}
b_i = 1 - \prod_{j \in \mathcal{N}_i} (1 - p \, c_{i \to j})  \; ,
\label{eq:bond1}
\end{equation}
and
\begin{equation}
\begin{array}{l}
c_{i \to j} = 1 - \prod_{k \in \mathcal{N}_{j} \setminus [
  \{i\} \cup (\mathcal{N}_j \cap \mathcal{N}_i)] } (1 - p c_{j \to k}) 
\\
\prod_{k \in \mathcal{N}_{j} \cap \mathcal{N}_{i}} [1 - p c_{j \to k} (1 - p + p c_{i \to k} )]  
\end{array} \; .
\label{eq:bond2}
\end{equation}
Here, $b_i$ and $c_{i \to j}$ have, in the bond percolation model, the
same meaning that $s_i$ and $t_{i \to j}$ have in site percolation.
The second equation explicitly imposes coherence of messages within
triangles. The message from node $k$ can in fact arrive to node $i$ in
two ways. (i) Along the path $k \to j \to i$ if the edge $(i, k)$ is
not active but the edge $(i, j)$ is active. This possibility happens
with probability $p c_{j \to k} (1 - p)$.  (ii) Simultaneously along
the paths $k \to j \to i$ and $k \to i$ if both edges $(i, k)$ and
$(i, j)$ are active. The latter possibility happens with probability
$p^2 c_{j \to k} c_{i \to k}$. In the absence of triangles, that means
$\mathcal{N}_{j} \cap \mathcal{N}_{i} = \emptyset$ for all edges
$(i,j)$, we recover the $M$-based approximation.  In the presence of
triangles instead, the additional correction term reduces the
estimated values of the variables $c$.  The system of
Eqs.~(\ref{eq:bond2}) can be solved by iteration. Its solutions can be
then plugged into Eqs.~(\ref{eq:bond1}), and the values of the
variables $b_i$ can finally be used to compute the bond percolation
strength as $B_\infty = N^{-1} \sum_i b_i$.
\begin{figure}[!htb]
  \begin{center}
    \includegraphics[width=0.47\textwidth]{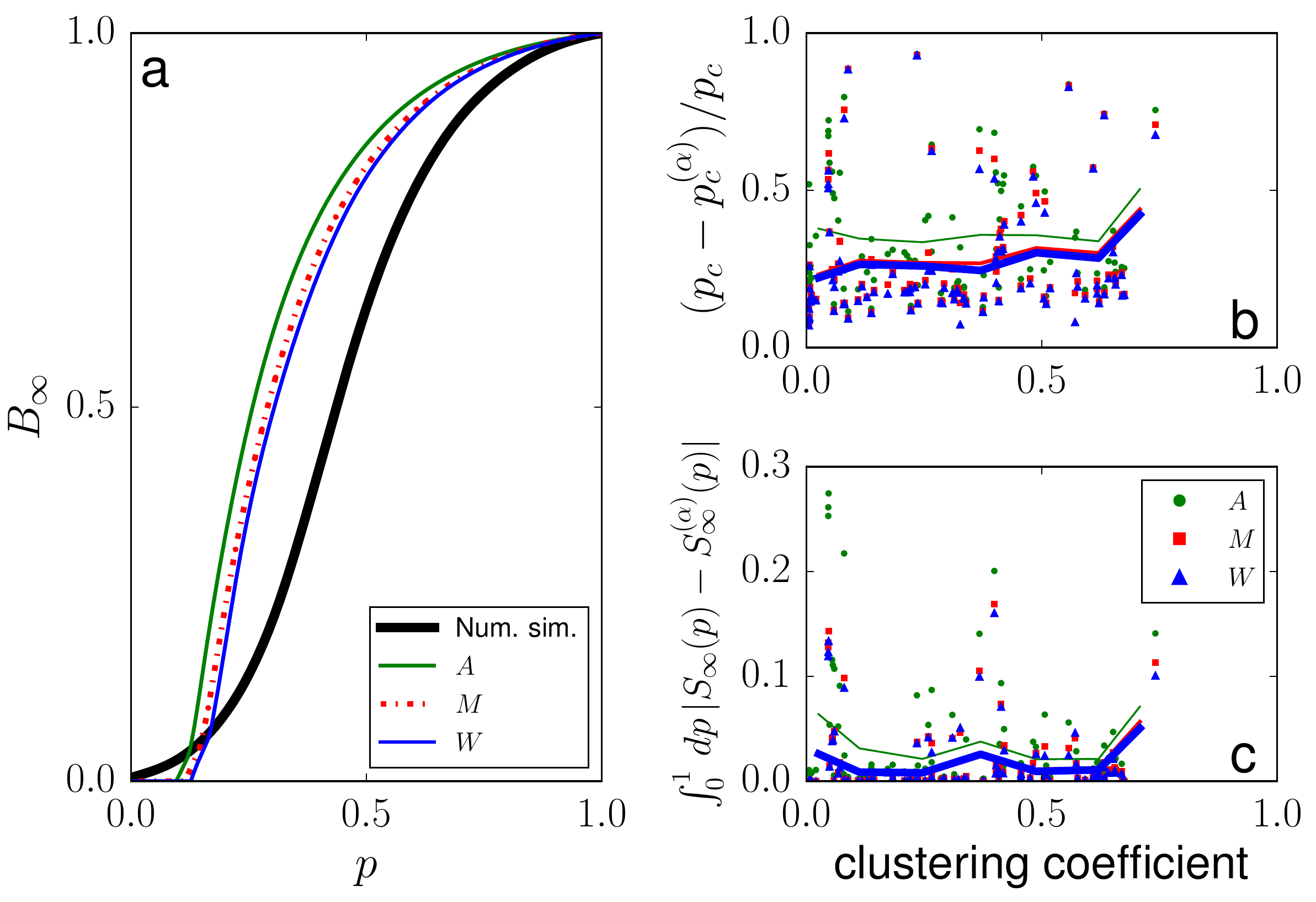}
  \end{center}
  \caption{(a) Phase diagram for the bond percolation
    model applied to co-authorship graph among
    network scientists~\cite{newman2006finding}. 
    The black line denotes the results
    of numerical simulations of the model. 
    The other curves
    represent results obtained through the numerical
    solution of the equations using different
    approximations. (b) Relative error of the various
    approximations in the estimation of the 
    bond percolation threshold. The figure
    represents the analogue of Fig.~\ref{fig2}a for
    bond percolation. (c) Same as Fig.~\ref{fig2}b but for
    bond percolation.
  }
\label{fig3}
\end{figure}
In Fig.~\ref{fig3}a, we compare the performance of the
approximations in reproducing the results of numerical
simulations in the same network analyzed in Fig.~\ref{fig1}b. 
The improvement in the prediction of the percolation
strength from the adjacency matrix-based up to the $W$-based 
approximation is not as significant as the one we found for site 
percolation.  The same
qualitative observation can be made for the other real networks we
analyzed (see SM). The linearization of the system of
Eqs.~(\ref{eq:bond2}) leads to the following vectorial equation for
the determination of the percolation threshold
\begin{equation}
\vec{c} = p_c \, W \vec{c} + p_c (1 - p_c) (M - W) \vec{c} \; .
\label{eq:pc_bond}
\end{equation}
The solution of this equation can be efficiently obtained by means of
a power-iteration algorithm combined with a binary search. As already
done for site percolation, we systematically test the
performance of the various approximations in $98$ real networks in
Figs.~\ref{fig3}b and~\ref{fig3}c. 
In general, accounting in this way for triangles improves only 
slightly the accuracy
of predictions with respect to the $M$-based approximation. 

In summary, our novel approximation goes, in a
relatively straightforward manner, beyond the locally tree-like ansatz.
The analysis carried out on real and synthetic networks allows to 
conclude that the $W$-based approximation greatly outperforms the 
$M$-based approximation for the site percolation process, leading
in almost all cases to an impressive agreement with numerical results.
For bond percolation instead the improvement is less
satisfactory and calls for further work.
Systematic approximations to account for
loops longer than three face severe intrinsic difficulties (see SM). 
It would be interesting to explore differences between the $M$-based
and the $W$-based approximations in the context of ordinary percolation
processes in interdependent networks~\cite{radicchi2015percolation} as
well in optimal percolation problems in isolated
ones~\cite{morone2015influence}.  As a final remark, we stress that the
improvement in the prediction of the percolation threshold comes at
a price.  Whereas the computational complexity of the algorithm is the
same in both $M$- and $W$-based approximations, the determination
of $p_c$ in the $W$-based approximation requires to deal with a
larger matrix. The Ihara-Bass determinant formula is able to
reduce the computation of the largest eigenvalue of the 
$2E \times 2E$ non-backtracking matrix $M$ to the largest eigenvalue of a 
$2N \times 2N$ matrix~\cite{bass1992ihara}.  
The quest for a similar formula for the matrix $W$
is an interesting challenge for future research.

\

This work is partially supported by the National Science Foundation
(Grant CMMI-1552487).


\begin{thebibliography}{27}
\expandafter\ifx\csname natexlab\endcsname\relax\def\natexlab#1{#1}\fi
\expandafter\ifx\csname bibnamefont\endcsname\relax
  \def\bibnamefont#1{#1}\fi
\expandafter\ifx\csname bibfnamefont\endcsname\relax
  \def\bibfnamefont#1{#1}\fi
\expandafter\ifx\csname citenamefont\endcsname\relax
  \def\citenamefont#1{#1}\fi
\expandafter\ifx\csname url\endcsname\relax
  \def\url#1{\texttt{#1}}\fi
\expandafter\ifx\csname urlprefix\endcsname\relax\def\urlprefix{URL }\fi
\providecommand{\bibinfo}[2]{#2}
\providecommand{\eprint}[2][]{\url{#2}}

\bibitem[{\citenamefont{Albert et~al.}(2000)\citenamefont{Albert, Jeong, and
  Barab{\'a}si}}]{albert2000error}
\bibinfo{author}{\bibfnamefont{R.}~\bibnamefont{Albert}},
  \bibinfo{author}{\bibfnamefont{H.}~\bibnamefont{Jeong}}, \bibnamefont{and}
  \bibinfo{author}{\bibfnamefont{A.-L.} \bibnamefont{Barab{\'a}si}},
  \bibinfo{journal}{Nature} \textbf{\bibinfo{volume}{406}},
  \bibinfo{pages}{378} (\bibinfo{year}{2000}).

\bibitem[{\citenamefont{Cohen et~al.}(2000)\citenamefont{Cohen, Erez,
  Ben-Avraham, and Havlin}}]{cohen2000resilience}
\bibinfo{author}{\bibfnamefont{R.}~\bibnamefont{Cohen}},
  \bibinfo{author}{\bibfnamefont{K.}~\bibnamefont{Erez}},
  \bibinfo{author}{\bibfnamefont{D.}~\bibnamefont{Ben-Avraham}},
  \bibnamefont{and} \bibinfo{author}{\bibfnamefont{S.}~\bibnamefont{Havlin}},
  \bibinfo{journal}{Phys. {R}ev. {L}ett.} \textbf{\bibinfo{volume}{85}},
  \bibinfo{pages}{4626} (\bibinfo{year}{2000}).

\bibitem[{\citenamefont{Callaway et~al.}(2000)\citenamefont{Callaway, Newman,
  Strogatz, and Watts}}]{callaway2000network}
\bibinfo{author}{\bibfnamefont{D.~S.} \bibnamefont{Callaway}},
  \bibinfo{author}{\bibfnamefont{M.~E.} \bibnamefont{Newman}},
  \bibinfo{author}{\bibfnamefont{S.~H.} \bibnamefont{Strogatz}},
  \bibnamefont{and} \bibinfo{author}{\bibfnamefont{D.~J.} \bibnamefont{Watts}},
  \bibinfo{journal}{Phys. {R}ev. {L}ett.} \textbf{\bibinfo{volume}{85}},
  \bibinfo{pages}{5468} (\bibinfo{year}{2000}).

\bibitem[{\citenamefont{Pastor-Satorras and
  Vespignani}(2001)}]{pastor2001epidemic}
\bibinfo{author}{\bibfnamefont{R.}~\bibnamefont{Pastor-Satorras}}
  \bibnamefont{and}
  \bibinfo{author}{\bibfnamefont{A.}~\bibnamefont{Vespignani}},
  \bibinfo{journal}{Phys. {R}ev. {L}ett.} \textbf{\bibinfo{volume}{86}},
  \bibinfo{pages}{3200} (\bibinfo{year}{2001}).

\bibitem[{\citenamefont{Newman}(2002)}]{newman2002spread}
\bibinfo{author}{\bibfnamefont{M.~E.} \bibnamefont{Newman}},
  \bibinfo{journal}{Phys. {R}ev. {E}} \textbf{\bibinfo{volume}{66}},
  \bibinfo{pages}{016128} (\bibinfo{year}{2002}).

\bibitem[{\citenamefont{Dorogovtsev et~al.}(2008)\citenamefont{Dorogovtsev,
  Goltsev, and Mendes}}]{dorogovtsev2008critical}
\bibinfo{author}{\bibfnamefont{S.~N.} \bibnamefont{Dorogovtsev}},
  \bibinfo{author}{\bibfnamefont{A.~V.} \bibnamefont{Goltsev}},
  \bibnamefont{and} \bibinfo{author}{\bibfnamefont{J.~F.}
  \bibnamefont{Mendes}}, \bibinfo{journal}{Rev. {M}od. {P}hys.}
  \textbf{\bibinfo{volume}{80}}, \bibinfo{pages}{1275} (\bibinfo{year}{2008}).

\bibitem[{\citenamefont{Dorogovtsev}(2010)}]{dorogovtsev2010lectures}
\bibinfo{author}{\bibfnamefont{S.~N.} \bibnamefont{Dorogovtsev}},
  \emph{\bibinfo{title}{Lectures on complex networks}},
  vol.~\bibinfo{volume}{24} (\bibinfo{publisher}{Oxford University Press
  Oxford}, \bibinfo{year}{2010}).

\bibitem[{\citenamefont{Newman}(2010)}]{newman2010networks}
\bibinfo{author}{\bibfnamefont{M.}~\bibnamefont{Newman}},
  \emph{\bibinfo{title}{Networks: an introduction}} (\bibinfo{publisher}{Oxford
  University Press}, \bibinfo{year}{2010}).

\bibitem[{\citenamefont{Cohen et~al.}(2002)\citenamefont{Cohen, Ben-Avraham,
  and Havlin}}]{cohen2002percolation}
\bibinfo{author}{\bibfnamefont{R.}~\bibnamefont{Cohen}},
  \bibinfo{author}{\bibfnamefont{D.}~\bibnamefont{Ben-Avraham}},
  \bibnamefont{and} \bibinfo{author}{\bibfnamefont{S.}~\bibnamefont{Havlin}},
  \bibinfo{journal}{Phys. {R}ev. {E}} \textbf{\bibinfo{volume}{66}},
  \bibinfo{pages}{036113} (\bibinfo{year}{2002}).

\bibitem[{\citenamefont{Serrano and
  Bogu\~n\'a}(2006{\natexlab{a}})}]{Serrano06}
\bibinfo{author}{\bibfnamefont{M.~A.} \bibnamefont{Serrano}} \bibnamefont{and}
  \bibinfo{author}{\bibfnamefont{M.}~\bibnamefont{Bogu\~n\'a}},
  \bibinfo{journal}{Phys. Rev. Lett.} \textbf{\bibinfo{volume}{97}},
  \bibinfo{pages}{088701} (\bibinfo{year}{2006}{\natexlab{a}}).

\bibitem[{\citenamefont{Serrano and
  Bogu\~n\'a}(2006{\natexlab{b}})}]{Serrano06b}
\bibinfo{author}{\bibfnamefont{M.~A.} \bibnamefont{Serrano}} \bibnamefont{and}
  \bibinfo{author}{\bibfnamefont{M.}~\bibnamefont{Bogu\~n\'a}},
  \bibinfo{journal}{Phys. Rev. E} \textbf{\bibinfo{volume}{74}},
  \bibinfo{pages}{056115} (\bibinfo{year}{2006}{\natexlab{b}}).

\bibitem[{\citenamefont{Gleeson}(2009)}]{Gleeson09}
\bibinfo{author}{\bibfnamefont{J.~P.} \bibnamefont{Gleeson}},
  \bibinfo{journal}{Phys. Rev. E} \textbf{\bibinfo{volume}{80}},
  \bibinfo{pages}{036107} (\bibinfo{year}{2009}).

\bibitem[{\citenamefont{Newman}(2009)}]{newman2009random}
\bibinfo{author}{\bibfnamefont{M.~E.} \bibnamefont{Newman}},
  \bibinfo{journal}{{P}hys. {R}ev. {L}ett.} \textbf{\bibinfo{volume}{103}},
  \bibinfo{pages}{058701} (\bibinfo{year}{2009}).

\bibitem[{\citenamefont{Miller}(2009)}]{Miller09}
\bibinfo{author}{\bibfnamefont{J.~C.} \bibnamefont{Miller}},
  \bibinfo{journal}{Phys. Rev. E} \textbf{\bibinfo{volume}{80}},
  \bibinfo{pages}{020901} (\bibinfo{year}{2009}).

\bibitem[{\citenamefont{Gleeson et~al.}(2010)\citenamefont{Gleeson, Melnik, and
  Hackett}}]{Gleeson10}
\bibinfo{author}{\bibfnamefont{J.~P.} \bibnamefont{Gleeson}},
  \bibinfo{author}{\bibfnamefont{S.}~\bibnamefont{Melnik}}, \bibnamefont{and}
  \bibinfo{author}{\bibfnamefont{A.}~\bibnamefont{Hackett}},
  \bibinfo{journal}{Phys. Rev. E} \textbf{\bibinfo{volume}{81}},
  \bibinfo{pages}{066114} (\bibinfo{year}{2010}).

\bibitem[{\citenamefont{Karrer et~al.}(2014)\citenamefont{Karrer, Newman, and
  Zdeborov\'a}}]{PhysRevLett.113.208702}
\bibinfo{author}{\bibfnamefont{B.}~\bibnamefont{Karrer}},
  \bibinfo{author}{\bibfnamefont{M.~E.~J.} \bibnamefont{Newman}},
  \bibnamefont{and}
  \bibinfo{author}{\bibfnamefont{L.}~\bibnamefont{Zdeborov\'a}},
  \bibinfo{journal}{Phys. {R}ev. {L}ett.} \textbf{\bibinfo{volume}{113}},
  \bibinfo{pages}{208702} (\bibinfo{year}{2014}).

\bibitem[{\citenamefont{Hashimoto}(1989)}]{hashimoto1989zeta}
\bibinfo{author}{\bibfnamefont{K.-i.} \bibnamefont{Hashimoto}},
  \bibinfo{journal}{Automorphic forms and geometry of arithmetic varieties.}
  pp. \bibinfo{pages}{211--280} (\bibinfo{year}{1989}).

\bibitem[{\citenamefont{Hamilton and Pryadko}(2014)}]{PhysRevLett.113.208701}
\bibinfo{author}{\bibfnamefont{K.~E.} \bibnamefont{Hamilton}} \bibnamefont{and}
  \bibinfo{author}{\bibfnamefont{L.~P.} \bibnamefont{Pryadko}},
  \bibinfo{journal}{Phys. {R}ev. {L}ett.} \textbf{\bibinfo{volume}{113}},
  \bibinfo{pages}{208701} (\bibinfo{year}{2014}).

\bibitem[{\citenamefont{Radicchi}(2015{\natexlab{a}})}]{radicchi2015percolation}
\bibinfo{author}{\bibfnamefont{F.}~\bibnamefont{Radicchi}},
  \bibinfo{journal}{Nature {P}hys.} \textbf{\bibinfo{volume}{11}},
  \bibinfo{pages}{597} (\bibinfo{year}{2015}{\natexlab{a}}).

\bibitem[{\citenamefont{Radicchi}(2015{\natexlab{b}})}]{radicchi2014predicting}
\bibinfo{author}{\bibfnamefont{F.}~\bibnamefont{Radicchi}},
  \bibinfo{journal}{Phys. Rev. E} \textbf{\bibinfo{volume}{91}},
  \bibinfo{pages}{010801} (\bibinfo{year}{2015}{\natexlab{b}}).

\bibitem[{\citenamefont{Faqeeh et~al.}(2015)\citenamefont{Faqeeh, Melnik, and
  Gleeson}}]{PhysRevE.91.052807}
\bibinfo{author}{\bibfnamefont{A.}~\bibnamefont{Faqeeh}},
  \bibinfo{author}{\bibfnamefont{S.}~\bibnamefont{Melnik}}, \bibnamefont{and}
  \bibinfo{author}{\bibfnamefont{J.~P.} \bibnamefont{Gleeson}},
  \bibinfo{journal}{Phys. Rev. E} \textbf{\bibinfo{volume}{91}},
  \bibinfo{pages}{052807} (\bibinfo{year}{2015}).

\bibitem[{\citenamefont{Newman}(2006)}]{newman2006finding}
\bibinfo{author}{\bibfnamefont{M.~E.} \bibnamefont{Newman}},
  \bibinfo{journal}{Phys. {R}ev. {E}} \textbf{\bibinfo{volume}{74}},
  \bibinfo{pages}{036104} (\bibinfo{year}{2006}).

\bibitem[{\citenamefont{Bollob{\'a}s et~al.}(2010)\citenamefont{Bollob{\'a}s,
  Borgs, Chayes, Riordan et~al.}}]{bollobas2010percolation}
\bibinfo{author}{\bibfnamefont{B.}~\bibnamefont{Bollob{\'a}s}},
  \bibinfo{author}{\bibfnamefont{C.}~\bibnamefont{Borgs}},
  \bibinfo{author}{\bibfnamefont{J.}~\bibnamefont{Chayes}},
  \bibinfo{author}{\bibfnamefont{O.}~\bibnamefont{Riordan}},
  \bibnamefont{et~al.}, \bibinfo{journal}{{A}nn. {P}robab.}
  \textbf{\bibinfo{volume}{38}}, \bibinfo{pages}{150} (\bibinfo{year}{2010}).

\bibitem[{\citenamefont{Melnik et~al.}(2011)\citenamefont{Melnik, Hackett,
  Porter, Mucha, and Gleeson}}]{Melnik11}
\bibinfo{author}{\bibfnamefont{S.}~\bibnamefont{Melnik}},
  \bibinfo{author}{\bibfnamefont{A.}~\bibnamefont{Hackett}},
  \bibinfo{author}{\bibfnamefont{M.~A.} \bibnamefont{Porter}},
  \bibinfo{author}{\bibfnamefont{P.~J.} \bibnamefont{Mucha}}, \bibnamefont{and}
  \bibinfo{author}{\bibfnamefont{J.~P.} \bibnamefont{Gleeson}},
  \bibinfo{journal}{Phys. Rev. E} \textbf{\bibinfo{volume}{83}},
  \bibinfo{pages}{036112} (\bibinfo{year}{2011}).

\bibitem[{\citenamefont{Radicchi and Castellano}(2015)}]{RadicchiCastellano15}
\bibinfo{author}{\bibfnamefont{F.}~\bibnamefont{Radicchi}} \bibnamefont{and}
  \bibinfo{author}{\bibfnamefont{C.}~\bibnamefont{Castellano}},
  \bibinfo{journal}{Nat. Commun.} \textbf{\bibinfo{volume}{6}},
  \bibinfo{pages}{10196} (\bibinfo{year}{2015}).

\bibitem[{\citenamefont{Morone and Makse}(2015)}]{morone2015influence}
\bibinfo{author}{\bibfnamefont{F.}~\bibnamefont{Morone}} \bibnamefont{and}
  \bibinfo{author}{\bibfnamefont{H.~A.} \bibnamefont{Makse}},
  \bibinfo{journal}{Nature} \textbf{\bibinfo{volume}{524}}, \bibinfo{pages}{65}
  (\bibinfo{year}{2015}).

\bibitem[{\citenamefont{Bass}(1992)}]{bass1992ihara}
\bibinfo{author}{\bibfnamefont{H.}~\bibnamefont{Bass}}, \bibinfo{journal}{Int.
  {J}. {M}ath.} \textbf{\bibinfo{volume}{3}}, \bibinfo{pages}{717}
  (\bibinfo{year}{1992}).

\end{thebibliography}

\end{document}